\documentclass[journal]{IEEEtran}

\ifCLASSINFOpdf
\else
   \usepackage[dvips]{graphicx}
\fi
\usepackage{url}

\usepackage{supertabular}
\usepackage{graphicx}
\usepackage{amsmath,empheq}
\newtheorem{theorem}{Theorem}
\usepackage{algorithm}
\usepackage{algorithmicx}
\usepackage{algpseudocode}
\usepackage{booktabs}
\usepackage{longtable}
\usepackage{mathtools}
\usepackage{amssymb}
\usepackage[table]{xcolor}
\mathtoolsset{showonlyrefs} 
\definecolor{light-gray}{HTML}{FFFFFF}
\definecolor{light-cyan}{HTML}{C4C4C4}
\usepackage{float}
\newcommand{\pluseq}{\mathrel{+}=}
\newcommand{\minuseq}{\mathrel{-}=}
\newcommand{\producteq}{\mathrel{*}=}
\usepackage{stfloats}
\usepackage{mdsymbol}
\usepackage{mathrsfs} 
\usepackage{mathtools}

\DeclarePairedDelimiter\floor{\lfloor}{\rfloor}
\DeclarePairedDelimiter\ceil{\lceil}{\rceil}

\begin{document}
\title{Hybrid Constructions of Binary Sequences with Low Autocorrelation Sidelobes}  

\author{Miroslav Dimitrov \IEEEmembership{Student Member, IEEE}, Tsonka Baitcheva \IEEEmembership{Member, IEEE}, and Nikolay Nikolov
\thanks{The research of the first and second author has been partially supported by the Bulgarian National Science Fund under contract number DH 12/8, 15.12.2017. The research of the third author was supported, in part, by a Bulgarian NSF contract KP-06-N32/2-2019.}
\thanks{M. Dimitrov  and T.  Baicheva are with the Institute of Mathematics and Informatics, Bulgarian Academy of Sciences, Sofia, Bulgaria (email:mirdim@math.bas.bg, email:tsonka@math.bas.bg).}
\thanks{N. Nikolov is with the State Agency of National Security, Sofia, Bulgaria.}}%

\maketitle

\begin{abstract}
In this work, a classical problem of the digital sequence design, or more precisely, finding binary sequences with optimal peak sidelobe level (PSL), is revisited. By combining some of our previous works, together with some mathematical insights, few hybrid heuristic algorithms were created. During our experiments, and by using the aforementioned algorithms, we were able to find PSL-optimal binary sequences for all those lengths, which were previously found during exhaustive searches by various papers throughout the literature. Then, by using  a general-purpose computer, we further demonstrate the effectiveness of the proposed algorithms by revealing binary sequences with lengths between 106 and 300, the majority of which possess record-breaking PSL values. Then, by using some well known algebraic constructions, we outline few strategies of finding highly-competitive binary sequences, which could be efficiently optimized, in terms of PSL, by the proposed algorithms.
\end{abstract}

\begin{IEEEkeywords}
Sequences, Peak Sidelobe Level (PSL), Hybrid Digital Sequence Design, Optimization
\end{IEEEkeywords}

\IEEEpeerreviewmaketitle

\section{Introduction}
\IEEEPARstart{D}igital sequence design plays an important role in various scientific domains, such as radar technology, telecommunications, active sensing systems, navigation, cryptography. One of the desirable characteristic a given binary sequence should posses is a low peak sidelobe level (\textbf{PSL}). Some well-known constructions of such sequences includes the Barker codes \cite{barker1953group}, Rudin-Shapiro sequences \cite{rudin1959some}\cite{shapiro1952extremal}, m-sequences  \cite{golomb1967shift}, Gold codes \cite{gold1967optimal}, Kasami codes \cite{kasami1966weight}, Weil sequences \cite{rushanan2006weil}, Legendre sequences \cite{pott2006finite}. Nevertheless, none of the aforementioned constructions guarantees that the generated binary sequence will posses the lowest possible (optimal) PSL value. Thus, currently,  initiating an exhaustive search is the only way to reveal an optimal PSL value for binary sequences of some fixed length. Given a binary sequence with length $n$, the PSL-optimal values for $n\leq 40$\cite{lindner1975binary}, $n\leq 48$\cite{baden1990optimal}, $n=64$\cite{coxson2005efficient}, $n\leq 68$\cite{leukhin2012binary}, $n\leq 74$\cite{leukhin2013optimal},  $n\leq 80$ \cite{leukhin2014exhaustive}, $n\leq 82$ \cite{leukhin2015bernasconi} and $n\leq 84$ \cite{leukhin2017exhaustive} are obtained. However, the PSL-optimal values of binary sequences with lengths $n$ greater than $84$ are still unknown. This is not surprising, since the search space of the set of all the binary sequences with some fixed length $n$ is $2^n$.

Since discovering a PSL-optimal value requires significant computational power, from practical point of view, the trade-off between optimality and complexity is justified, i.e. the usage of an algorithm having significantly lower complexity compared to the exhaustive search routine, which is capable of reaching candidates close to the PSL-optimal ones (near-optimal).  The state-of-the-art strategies for near-optimal PSL binary sequence construction include CAN \cite{he2009designing}, ITROX \cite{soltanalian2012computational}, MWISL-Diag, MM-PSL \cite{song2015sequence}, DPM \cite{kerahroodi2017coordinate}, 1bCAN \cite{lin2019efficient}, shotgun hill climbing (SHC) \cite{dimitrov2020efficient} and optimized for long binary sequences hill climbing (HC) \cite{dimitrov2020generation}. 

The currently known PSL records for $ 85 \leq n \leq 105$ are published in \cite{nunn2008best}, and for $ 106 \leq n \leq 300$ in \cite{dimitrov2020efficient}\cite{dzvonkovskaya2008long}\cite{Patent}\cite{mow2015new}\cite{coxson2020long}. Furthermore, some of the aforementioned works published records for some chosen  lengths of $n \geq 300$ as well. For example, in \cite{coxson2014adiabatic} a D-Wave 2 quantum computer was used, altogether with an adiabatic quantum algorithm, for searching binary sequences with low PSL up to lengths of 426. 

In Section \ref{sec:problemRevisited}, we demonstrate that the hybridization of two distinct PSL-optimizing algorithms could be beneficial to the overall goal of finding near-optimal PSL binary sequences. In fact, during our experiments in Sections \ref{sec:fitness} and \ref{sec:resultsMain}, and by using just a general purpose processor, we were able to find PSL-optimal binary sequences for all those lengths, which were previously discovered by an exhaustive search only. Then, by using the latest hybrid strategy, record-breaking PSL values for almost all binary sequences with lengths in $\left[106,300\right]$ are revealed. Finally, in Section \ref{sec:hybrid}, we investigate the applicability of the proposed algorithm as an extension to some well-known algebraic constructions.

\section{Preliminaries}
\label{sec:prelims}

We denote as $B=(b_0,b_1,\cdots ,b_{n-1})$ the binary sequence with length $n>1$, such that $b_i\in \{-1,1\}, 0\leq i\leq n-1$. The aperiodic autocorrelation function of $B$ is given by $$C_u(B)=\sum_{j=0}^{n-u-1} b_jb_{j+u}, \ \ for \ u\in \{0,1,\cdots, n-1\}.$$

We define $C_u(B)$ for $u\in \{1, \cdots ,n-1\}$ as a sidelobe level. $C_0(B)$ is called the mainlobe. We define the PSL of $B$ as $$B_{PSL}=\max_{0<u<n} \lvert C_u(B)\rvert.$$

An m-sequence $M=(x_0, x_1, \cdots, x_{2^m-2})$ of length $2^m-1$ is defined by: $$x_i = {\left(-1\right)}^{Tr(\beta \alpha^i)}, \text{ for } 0 \leq i < 2^m-1,$$ where $\alpha$ is a primitive element of the field $\mathbb{F}_{2^m}$, $\beta \in \mathbb{F}_{2^m}$, and $Tr$ is denoting the trace function from $\mathbb{F}_{2^m}$ to $\mathbb{F}_2$.

Given an odd prime $p$, a Legendre sequence L with  length p is defined by:
 \begin{equation*}
  L_i = \begin{cases}
        1 \text{, if $i$ is a quadratic residue mod $p$}
        \\
        -1 \text{, otherwise}.
        \end{cases}
 \end{equation*}
 
We denote as $B \leftarrow \rho$ the binary sequence obtained from $B$, by left-rotating it $\rho$ times. By definition, $B \leftarrow \lvert B \rvert \equiv B$. Furthermore, if $b_i$ is the element of $B$ on position $i$, we will denote as $b_i^{\leftarrow \rho}$ the element of $B \leftarrow \rho$ on position $i$.

Let us denote $C_{n-i-1}(B)$ by $\hat{C}_i(B)$. Since this is just a rearrangement of the sidelobes of $B$, it follows that:
\[
	B_{PSL}=\max_{0<u<n} \lvert C_u(B)\rvert = \max_{0 \le u < n-1} \lvert \hat{C}_u(B)\rvert.
\]

\section{PSL problem revisited}
\label{sec:problemRevisited}

Throughout this section, a brief overview of the existing PSL-optimizing algorithms was made. In \cite{dimitrov2020generation} a comparison of the state-of-the-art algorithms, in terms of algorithm efficiency (the ratio of the beneficial work performed by the algorithm to the total energy invested) and actual effectiveness (the quality of the achieved results) was made. The best results were achieved by \textbf{SHC} \cite{dimitrov2020efficient} algorithm, regarding the binary sequences with length less than $300$, and \textbf{HC} \cite{dimitrov2020generation}, for all the remaining lengths. However, the approximated binary sequence's length, from which HC starts outperforming SHC, is fuzzy and yet to be determined.

In Table \ref{tab:comparisons} a comparison between the most significant components of SHC and HC was made. In summary, both heuristic algorithms are not deterministic, i.e. starting from two identical states rarely  results in two identical ending states. The search operator used in both SHC and HC is the single flip operator. Thus, each modification is a simple composition of single flips. One major difference between the two algorithms is their complexity. Indeed, in HC the time complexity of the flip operation is linear, which is significant advantage compared to the quadratic one to be found in SHC. Another major difference between HC and SHC is the probability of missing (fails to detect) a better binary sequence, which is just 1 flip away from the current position.

As observed in \cite{dimitrov2020generation}, the PSL-optimization process of very long binary sequences is a time consuming routine, despite the algorithm linear time and memory complexities. Thus, HC avoids restarts, i.e. re-initializing the starting state with a pseudo-random binary sequence. However, re-initialization appears to be significantly beneficial when dealing with PSL-optimization of binary sequences with relatively small lengths, such as the SHC algorithm.

By considering the observations made above, we have revisited the SHC algorithm:

\begin{itemize}
\item{The quadratic flip operator was interchanged with the linear flip operator}
\item{The probing strategy (searching for better candidates) was interchanged with the more efficient probing strategy introduced in HC}
\end{itemize}

For convenience, the flip operator from \cite{dimitrov2020generation} is given in Algorithm \ref{algor:InMemoryFlip}. The function \textbf{Flip} takes three parameters as input:

\begin{itemize}
\item{$f$ - bit position to be flipped}
\item{$\Psi$ - binary sequence as an array}
\item{$\Omega_{\Psi}$ - the sidelobes of $\Psi$ as an array}
\end{itemize}

\algrenewcommand\algorithmicindent{0.5em}%
\begin{algorithm}[]
\caption{The in-memory flip introduced in \cite{dimitrov2020generation}}
\label{algor:InMemoryFlip}
\begin{algorithmic}[1]
\Procedure{Flip}{$f, \Psi, \Omega_{\Psi}$}
\State $n=\lvert\Psi\rvert$
\State $\delta_{min} \gets \min_{\left(n-f-1, f\right)}$
\State $\delta_{max} \gets \max_{\left(n-f, f \right)}$
\If{$f \leq \frac{n-1}{2}$}
	\For{$q \in \left[0, \delta_{max}-\delta_{min}-1\right)$}
	\State $\Omega_{\Psi}[\delta_{min}+q] \minuseq 2\Psi[f]\Psi[n-q-1]$
	\EndFor
\Else
	\For{$q \in \left[0, \delta_{max}-\delta_{min}\right)$}
	\State $\Omega_{\Psi}[\delta_{min}+q] \minuseq 2\Psi[f]\Psi[q]$
	\EndFor	
\EndIf

\If{$f \leq \frac{n-1}{2}$}
	\For{$q \in \left[0, n-\delta_{max}\right)$}
	\State $\Omega_{\Psi}[\delta_{max}+q-1] \minuseq 2\Psi[f]\left(\Psi[2f-q]+\Psi[q]\right)$
	\EndFor
\Else
	\For{$q \in \left[0, n-\delta_{max}-1\right)$}
		\State {$\Omega_{\Psi}[\delta_{max}+q] \minuseq$} \State {$2\Psi[f]\left(\Psi[\delta_{max}-\delta_{min}+q]+\Psi[n-q-1]\right)$}
	\EndFor
\EndIf
\State $\Psi[f] \producteq -1 $
\EndProcedure
\end{algorithmic}
\end{algorithm} 

The complete pseudo-code of the kernel of the revisited SHC algorithm is summarized in Algorithm \ref{algor:SHCkernel}. For brevity, the following notations were used:

\begin{itemize}
\item{$n$ - the binary sequence's length}
\item{$\mathbb{T}$ - the threshold value of the instance}
\item{$F$ - a fixed fitness function}
\item{$V$, $V^*$ - respectively the current best and the overall best fitness value}
\item{$c$ - the counter. The algorithm quits if the counter $c$ reaches the threshold $\mathbb{T}$ }
\item{$\mathbb{Z}^+_n$} - the set of all positive integer numbers strictly less than $n$ 
\item{$\mathbb{L}$, $\mathbb{G}$ - binary variables: $\mathbb{L}$ (local) is activated if $V$ is improved, while $\mathbb{G}$ (global) is activated if $V^*$ is improved}
\item{$\mathbb{B}^n$} - the set of all $n$-dimensional binary sequences with elements from $\lbrace -1,1 \rbrace$ 
\item{$Q$ - the quaking function as defined in \cite{dimitrov2020generation}. For example, if the input triplet of $Q$ is $x,L,SL$, the function flips $x$ random bits in $L$, and at the same time, in-memory updating the sidelobe array $SL$ }
\end{itemize}

\algrenewcommand\algorithmicindent{0.5em}%
\begin{algorithm}[]
\caption{The SHC revisited kernel}
\label{algor:SHCkernel}
\begin{algorithmic}[1]
\Procedure{SHC}{$n, \mathbb{T}$}

\State pick $\Psi \in \mathbb{B}^n$ 
\State {$V^*$, $V$, $\mathbb{G}$, $\mathbb{L}$, $c$ $\gets $ $F(\Omega_{\Psi})$, 0, True, False, 0}
\While {$c < \mathbb{T}$}
	\State $c \pluseq 1$
	\If{$\mathbb{G}$}
		\State pick $r \in \mathbb{Z}^+_n$
		\For{$i \in \left[0, n\right)$}
			\State flip$\left(\left(r+i\right)\%n, \Psi, \Omega_{\Psi}\right)$
			\If{$V^* >  F(\Omega_{\Psi})$}
				\State $V^*$, $\mathbb{L}$ $\gets F(\Omega_{\Psi})$, True				
				\State \textbf{break}
			\Else
				\State flip$\left(\left(r+i\right)\%n, \Psi, \Omega_{\Psi}\right)$
			\EndIf
		\EndFor
		\If {$\mathbb{L}$}
			\State $\mathbb{G}$, $\mathbb{L}$ $\gets$ True, False
			\State \textbf{continue}
		\Else
			\State $\mathbb{G}$ $\gets$ False
		\EndIf
	\Else
		\State pick r $\in \mathbb{Z}^+_4$
		\State $Q$(1+r, $\Psi$, $\Omega_{\Psi}$)
		\State $\mathbb{G}$, $\mathbb{L}$ $\gets$ True, False
	\EndIf
\EndWhile
\EndProcedure
\end{algorithmic}
\end{algorithm}  

\begin{table}
\begin{center}
\caption{A comparison between SHC and HC}
\label{tab:comparisons}
\ttfamily
\rowcolors{2}{light-gray}{light-cyan}
\begin{tabular}{lllll}
& {SHC} & {HC} \\
\toprule
\textbf{Deterministic} & No & No\\
\textbf{Search Operator} & Flip & Flip\\
\textbf{Complexity} & $O(n^2)$ & $O(n)$\\
\textbf{Fitness Function} & $x^4$ & $x^4$ \\
\textbf{Restarts} & Yes & No \\
\textbf{Missing Probability} & $>0$ & $=0$ \\
\showrowcolors
\bottomrule
\end{tabular}
\end{center}
\end{table}

\section{Fitness Functions}
\label{sec:fitness}

In this section, considering the significant changes made in the SHC algorithm, the fitness functions parameters are carefully analyzed, re-evaluated and updated. Given a binary sequence $\Psi$, both algorithms (SHC and HC) are sharing the same fitness function $F$, s.t: \[ F(\Psi) = \sum_{x \in \Omega_{\Psi}}{\mathopen| x \mathclose| ^4} = \sum_{x \in \Omega_{\Psi}}{x^4} \] During our previous experiments, we reached to the conclusion that interchanging the power $4$ with larger or smaller value, is respectively too intolerant or too tolerant to the largest elements in $\Omega_{\Psi}$. However, since significant changes to the kernel of SHC were made, this observation is to be re-evaluated by series of experiments. More precisely, given a fixed threshold $\mathbb{T}$, and the fitness function $\sum_{x \in \Omega_{\Psi}}{\mathopen| x \mathclose| ^\alpha}$, a comparison between the efficiency of different $\alpha$ values is measured.

In Table \ref{tab:alfa100} the results regarding binary sequences with length 100 are given. Each row of the table corresponds to a different experiment. For a more informative measurement of the overall efficiency of the experiments, another variable $V^{\triangledown}$ was introduced. It measures the median value of all the best values $V^{*}$. More formally, if $t_i$ denotes the thread $i$ of a given experiment $\mathbb{E}$ with $\mathbb{R}$ restarts, and if the best results achieved by $t_i$ is denoted as $V^{*}_i$, then $$V^{\triangledown} = \frac{\sum_{i \in \mathbb{E}}{V^*_i}}{\mathbb{R}}$$

At first, the numerical experiments suggest $\alpha=3$ as a near-optimal value for achieving best results. Indeed, given a binary sequence with length 100, and $(\alpha, \mathbb{R}, \mathbb{T}) = (3, {10}^2, {10}^4)$, the  value of $V^{\triangledown}$ is smaller compared to the other experiments' values. This observation is more clearly visible throughout the experiments with binary sequences having length 256 summarized in Table \ref{tab:alfa256} and binary sequences with length 500 (see Table \ref{tab:alfa500} and the triplet $(\alpha, \mathbb{R}, \mathbb{T}) = (3, {10}^2, {10}^4)$ with $V^{\triangledown} = 11.51$). However, this tendency of $\alpha=3$ supremacy over integer values of $\alpha$ is not observable throughout larger values of $n$. As summarized in Table \ref{tab:alfa1024}, the triplet $(\alpha, \mathbb{R}, \mathbb{T}) = (4, {10}^2, {10}^4)$ yields better characteristics than $(\alpha, \mathbb{R}, \mathbb{T}) = (3, {10}^2, {10}^4)$. In fact, the quality of the binary sequences yielded by the triplet $(\alpha, \mathbb{R}, \mathbb{T}) = (4, {10}^2, {10}^3)$, having $V^{\triangledown}$ equal to $24.81$, is almost the same as those binary sequences generated by the triplet $(\alpha, \mathbb{R}, \mathbb{T}) = (3, {10}^2, {10}^4)$ with $V^{\triangledown}=24.98$. Since the first threshold value (${10}^3$) is ten times smaller than the second one (${10}^4$), and given the negligible difference of the binary sequences' quality ($0.17$), this correlation is particularly beneficial and could be further exploited to reduce the overall time needed for the binary sequences optimization routines.

\begin{table}
\begin{center}
\caption{Efficiency and comparison of various triplets $\left( \alpha, \mathbb{T}, 100 \right)$}
\label{tab:alfa100}
\ttfamily
\rowcolors{2}{light-gray}{light-cyan}
\begin{tabular}{llllll}
$n$ & $\alpha$ & $\mathbb{R}$ & $\mathbb{T}$ & $V^*$ & $V^{\triangledown}$ \\
\toprule
100 & 1 & $10^2$ & $10^3$ & $7$ & $7.63$\\
100 & 1 & $10^2$ & $10^4$ & $6$ & $7.00$\\
100 & 2 & $10^2$ & $10^3$ & $6$ & $6.95$\\
100 & 2 & $10^2$ & $10^4$ & $6$ & $6.72$\\
100 & 3 & $10^2$ & $10^3$ & $6$ & $6.94$\\
100 & 3 & $10^2$ & $10^4$ & $6$ & $6.70$\\
100 & 4 & $10^2$ & $10^3$ & $7$ & $7.00$\\
100 & 4 & $10^2$ & $10^4$ & $6$ & $6.94$\\
100 & 5 & $10^2$ & $10^3$ & $7$ & $7.00$\\
100 & 5 & $10^2$ & $10^4$ & $6$ & $6.95$\\
100 & 6 & $10^2$ & $10^3$ & $7$ & $7.10$\\
100 & 6 & $10^2$ & $10^4$ & $7$ & $7.00$\\
100 & 7 & $10^2$ & $10^3$ & $7$ & $7.23$\\
100 & 8 & $10^2$ & $10^3$ & $8$ & $8.26$\\

\showrowcolors
\bottomrule
\end{tabular}
\end{center}
\end{table}

\begin{table}
\begin{center}
\caption{Efficiency and comparison of various triplets $\left( \alpha, \mathbb{T}, 256 \right)$}
\label{tab:alfa256}
\ttfamily
\rowcolors{2}{light-gray}{light-cyan}
\begin{tabular}{llllll}
$n$ & $\alpha$ & $\mathbb{R}$ & $\mathbb{T}$ & $V^*$ & $V^{\triangledown}$ \\
\toprule
256 & 1 & $10^2$ & $10^3$ & $13$ & $14.66$\\
256 & 1 & $10^2$ & $10^4$ & $13$ & $13.94$\\
256 & 2 & $10^2$ & $10^3$ & $11$ & $11.98$\\
256 & 2 & $10^2$ & $10^4$ & $11$ & $11.72$\\
256 & 3 & $10^2$ & $10^3$ & $11$ & $11.92$\\
256 & 3 & $10^2$ & $10^4$ & $11$ & $11.51$\\
256 & 4 & $10^2$ & $10^3$ & $11$ & $11.99$\\
256 & 4 & $10^2$ & $10^4$ & $11$ & $11.84$\\
256 & 5 & $10^2$ & $10^3$ & $12$ & $12.22$\\

\showrowcolors
\bottomrule
\end{tabular}
\end{center}
\end{table}

\begin{table}
\begin{center}
\caption{Efficiency and comparison of various triplets $\left( \alpha, \mathbb{T}, 500 \right)$}
\label{tab:alfa500}
\ttfamily
\rowcolors{2}{light-gray}{light-cyan}
\begin{tabular}{llllll}
$n$ & $\alpha$ & $\mathbb{R}$ & $\mathbb{T}$ & $V^*$ & $V^{\triangledown}$ \\
\toprule
500 & 1 & $10^2$ & $10^3$ & $21$ & $23.19$\\
500 & 1 & $10^2$ & $10^4$ & $21$ & $22.10$\\
500 & 2 & $10^2$ & $10^3$ & $17$ & $17.83$\\
500 & 2 & $10^2$ & $10^4$ & $16$ & $17.04$\\
500 & 3 & $10^2$ & $10^3$ & $16$ & $16.94$\\
500 & 3 & $10^2$ & $10^4$ & $16$ & $16.61$\\
500 & 4 & $10^2$ & $10^3$ & $16$ & $17.04$\\
500 & 4 & $10^2$ & $10^4$ & $16$ & $16.89$\\

\showrowcolors
\bottomrule
\end{tabular}
\end{center}
\end{table}

\begin{table}
\begin{center}
\caption{Efficiency and comparison of various triplets $\left( \alpha, \mathbb{T}, 1024 \right)$}
\label{tab:alfa1024}
\ttfamily
\rowcolors{2}{light-gray}{light-cyan}
\begin{tabular}{llllll}
$n$ & $\alpha$ & $\mathbb{R}$ & $\mathbb{T}$ & $V^*$ & $V^{\triangledown}$ \\
\toprule
1024 & 1 & $10^2$ & $10^3$ & $34$ & $38.50$\\
1024 & 1 & $10^2$ & $10^4$ & $34$ & $35.96$\\
1024 & 2 & $10^2$ & $10^3$ & $27$ & $28.27$\\
1024 & 2 & $10^2$ & $10^4$ & $26$ & $27.12$\\
1024 & 3 & $10^2$ & $10^3$ & $24$ & $25.43$\\
1024 & 3 & $10^2$ & $10^4$ & $24$ & $24.81$\\
1024 & 4 & $10^2$ & $10^3$ & $24$ & $24.98$\\
1024 & 4 & $10^2$ & $10^4$ & $24$ & $24.16$\\
1024 & 5 & $10^2$ & $10^3$ & $25$ & $25.32$\\
1024 & 6 & $10^2$ & $10^3$ & $25$ & $25.98$\\

\showrowcolors
\bottomrule
\end{tabular}
\end{center}
\end{table}

During the final two experiments, considering the bigger sizes of the binary sequences, the threshold value is fixed to $10^3$. However, the data gathered throughout the previous experiments suggested that if we have a triplet $(n, \mathbb{R}, \mathbb{T}_1)$ measured with $V^{\triangledown}_1$, then, given $\mathbb{T}_1 \geq {10^3}$ and some threshold value $\mathbb{T}_2 >> \mathbb{T}_1$, such that the triplet $(n, \mathbb{R}, \mathbb{T}_2)$ is measured with $V^{\triangledown}_2$, then $V^{\triangledown}_2 < V^{\triangledown}_1$. 

In Tables \ref{tab:alfa2048} and \ref{tab:alfa4096}, triplets of the form $(\alpha, {10}^2, {10}^3)$ were analyzed, corresponding to binary sequences with respective lengths of 2048 and 4096. It appears that the longer the binary sequences is ($n$), the larger the aggression of the optimization routine should be ($\alpha$). Indeed, in the case of $n=2048$, the best value of $V^{\triangledown}=36.74$ is calculated by using $\alpha=5$, while in the case of binary sequences with lengths $n=4096$, the best value of $V^{\triangledown}=54.16$ is yielded by using $\alpha=6$.

\begin{table}
\begin{center}
\caption{Efficiency and comparison of various triplets $\left( \alpha, \mathbb{T}, 2048 \right)$}
\label{tab:alfa2048}
\ttfamily
\rowcolors{2}{light-gray}{light-cyan}
\begin{tabular}{llllll}
$n$ & $\alpha$ & $\mathbb{R}$ & $\mathbb{T}$ & $V^*$ & $V^{\triangledown}$ \\
\toprule
2048 & 1 & $10^2$ & $10^3$ & $58$ & $65.64$\\
2048 & 2 & $10^2$ & $10^3$ & $41$ & $44.32$\\
2048 & 3 & $10^2$ & $10^3$ & $37$ & $38.27$\\
2048 & 4 & $10^2$ & $10^3$ & $36$ & $36.99$\\
2048 & 5 & $10^2$ & $10^3$ & $36$ & $36.74$\\
2048 & 6 & $10^2$ & $10^3$ & $36$ & $36.91$\\

\showrowcolors
\bottomrule
\end{tabular}
\end{center}
\end{table}

\begin{table}
\begin{center}
\caption{Efficiency and comparison of various triplets $\left( \alpha, \mathbb{T}, 2048 \right)$}
\label{tab:alfa4096}
\ttfamily
\rowcolors{2}{light-gray}{light-cyan}
\begin{tabular}{llllll}
$n$ & $\alpha$ & $\mathbb{R}$ & $\mathbb{T}$ & $V^*$ & $V^{\triangledown}$ \\
\toprule
4096 & 1 & $10^2$ & $10^3$ & $99$ & $110.11$\\
4096 & 2 & $10^2$ & $10^3$ & $64$ & $68.48$\\
4096 & 3 & $10^2$ & $10^3$ & $55$ & $57.47$\\
4096 & 4 & $10^2$ & $10^3$ & $53$ & $54.91$\\
4096 & 5 & $10^2$ & $10^3$ & $53$ & $54.17$\\
4096 & 6 & $10^2$ & $10^3$ & $53$ & $54.16$\\
4096 & 7 & $10^2$ & $10^3$ & $53$ & $54.28$\\

\showrowcolors
\bottomrule
\end{tabular}
\end{center}
\end{table}

\section{Practical Applications and Results}
\label{sec:resultsMain}

The observations made throughout the experiments in Section \ref{sec:fitness}, as well as the more efficient algorithm constructed in Section \ref{sec:problemRevisited}, motivated us to revisit the PSL-optimization problem.

\subsection{Finding PSL-optimal binary sequences heuristically}
\label{subsec:1}

As previously discussed, binary sequences with lengths up to $84$ and PSL-optimal values have been already discovered by using various exhaustive search strategies.    This data is particularly beneficial for measuring the efficiency of a given PSL-optimizing algorithm. In other words, given a search space with binary sequences with some fixed length $n \leq 84$, and some PSL-optimizing algorithm $\mathbb{A}$ with a reasonable threshold value,   the best results achieved by $\mathbb{A}$ could be compared with the already known optimal PSL values. 

During our experiments, we have used a single general purpose computer with a 6-cored central processing unit architecture, capable of running 12 threads simultaneously. Surprisingly, by using the SHC revisited kernel, as well as fixed value of $\alpha=2$, we were able to reach binary sequences with optimal PSL values for each length in $\left[1,82\right]$. Given the linear time and memory complexities of the algorithm, for the majority of those lengths, the PSL-optimal binary sequences were reached for less than a minute. However, for some border cases, the needed time was few hours. The best results yielded by our experiments are summarized in Table \ref{tab:ReachedOptimalSolutions}. A remark should be made, that we have included just one PSL-optimal binary sequence for a given length. However, for almost each fixed length, the algorithm was able to find more than one binary sequences having an optimal PSL value. The binary sequences are given in a hexadecimal format, by omitting the leading zeroes. In the last column of Table \ref{tab:ReachedOptimalSolutions}, beside the corresponding optimal PSL value of the hexadecimal binary sequence given in column 2, the symbol $\hourglass$ was used to illustrate some approximation of the time needed for Algorithm 2 to reach a PSL-optimal binary sequence:
\begin{itemize}
\item{$\hourglass \approx$ minute}
\item{$\hourglass\hourglass \approx$ hour}
\item{$\hourglass\hourglass\hourglass \approx$ day}
\end{itemize}
For all other cases, the algorithm was able to reach the optimal PSL for less than a minute, and in some cases, for less than a second.

\begin{table}
\begin{center}
\caption{Reached optimal solutions}
\label{tab:ReachedOptimalSolutions}
\ttfamily
\rowcolors{2}{light-gray}{light-cyan}
\begin{tabular}{lll}
$n$ & Sequence in HEX & PSL \\
\toprule
10 & 37a & 3 \\
11 & 712 & 1 \\
12 & b3 & 2 \\
13 & a60 & 1 \\
14 & 2a60 & 2 \\
15 & 3dba & 2 \\
16 & a447 & 2 \\
17 & 1c0a6 & 2 \\
18 & 2650f & 2 \\
19 & 52447 & 2 \\
20 & 87b75 & 2 \\
21 & 129107 & 2 \\
22 & 14f668 & 3 \\
23 & 56ce01 & 3 \\
24 & 4a223c & 3 \\
25 & 9b501c & 2 \\
26 & 2e7e935 & 3 \\
27 & 24bb9f1 & 3 \\
28 & e702a49 & 2 \\
29 & 10e2225b & 3 \\
30 & 2a31240f & 3 \\
31 & 2d079910 & 3 \\
32 & 2857d373 & 3 \\
33 & 16915cc18 & 3 \\
34 & 1a43808dd & 3 \\
35 & 5569e0199 & 3 \\
36 & 87885776d & 3 \\
37 & 10c1237a2b & 3 \\
38 & 7caacc212 & 3 \\
39 & 29ca6c7c80 & 3 \\
40 & 22471e86fa & 3 \\
41 & 7c64d77ade & 3 \\
42 & 4447b874b4 & 3 \\
43 & 550e7f99b49 & 3 \\
44 & cb4b8778888 & 3 \\
45 & b6cab731e3f & 3 \\
46 & 16959a2e3003 & 3 \\
47 & 69a7e851988 & 3 \\
48 & e6e9bd5bc10f & 3 \\
49 & 103f6eda6ae71 & 4 \\
50 & 31dceade9920f & 4 \\
51 & 71c077376adb4 & 3 \\
52 & 600dc3cb4cd56 & 4 \\
53 & 1671848a940fcb & 4 \\
54 & 2622a797806912 & 4 \\
55 & 6006a578ea6933 & 4 \\
56 & 61e4b3229420af & 4 \\
57 & 143606103beca35 & 4 \\
58 & 215081f5644f2ce & 4 \\
59 & 3b06774134bdf5e & 4 \\
60 & 4df905215263a39 & 4 \\
61 & 193c99e12d6010aa & 4 \\
62 & 25695564e679ff83 & 4 \\
63 & 707d54b9c99ef690 & 4 \\
64 & d4ef33d372e082be & 4 \\
65 & 1f75f6c8f84c6b50 & 4 \\
66 & 28a59401e57b1c993 & 4 \\
67 & 5ba4d417723078421 & 4 \\
68 & d155a49d98c7bf7e1 & 4 \\
69 & 18ff3eb05d654b6665 & 4 \\
70 & 2b5aae6765e79b600f & 4 \\
71 & 8cea0ff5e92cb9726 & 4 \\
72 & dbcf036102615ab2a & 4$^\hourglass$ \\
73 & 164da9aab5398f1ffe1 & 4$^\hourglass$ \\
74 & 8c9c6dab51e57580f & 4$^\hourglass$ \\
75 & 5ff692ba8d62f1e3326 & 4 \\
76 & 87ad414fa9fcbb99a6c & 4 \\
77 & fe00861c0d932958aca & 4 \\
78 & 328b457f0461e4ed7b73 & 4$^{\hourglass\hourglass}$ \\
79 & 55fae4fdb42732de2ce2 & 4$^\hourglass$ \\
80 & fe00a22a539352e3659e & 4 \\
81 & dc9df3ff085a6c3aae53 & 4$^{\hourglass\hourglass}$ \\
82 & 2bf0fceee2499527bc61a & 4$^{\hourglass\hourglass\hourglass}$ \\
\showrowcolors
\bottomrule
\end{tabular}
\end{center}
\end{table}

\subsection{Finding PSL-near-optimal binary sequences heuristically}
\label{subsec:2}

In \cite{leukhin2017exhaustive} it was shown that there are no binary sequences with lengths 83 or 84 with PSL 4 (or less). For completeness, in Table \ref{tab:ReachedNearOptimalSolutionsPartI}, two binary sequences (with lengths 83 and 84) reached by Algorithm 2 are given. Both possess an optimal PSL value and were reached for less than a minute.

The near-optimal PSL values for binary sequences with lengths from 85 to 105 are found in \cite{nunn2008best}. However, there is no further information regarding the particular optimization technique that was applied. The authors just stated that \textit{`The searches involved a combination of several global optimization methods`}. Hence, it is difficult to recreate the experiment or, for example, apply the aforementioned mix of unknown optimization techniques to binary sequences with different (greater) lengths. Nevertheless, by using Algorithm 2, we were able to reach the same PSL values for all the binary sequences with lengths from 85 to 105 (see Table \ref{tab:ReachedNearOptimalSolutionsPartII}). It should be mentioned that the binary sequences from Table \ref{tab:ReachedNearOptimalSolutionsPartII}) are different from those that were previously published in the literature. 

\begin{table}
\begin{center}
\caption{Reached optimal solutions - continued}
\label{tab:ReachedNearOptimalSolutionsPartI}
\ttfamily
\rowcolors{2}{light-gray}{light-cyan}
\begin{tabular}{lll}
$n$ & Sequence in HEX & PSL \\
\toprule
83 & 7fc3af0a919735c4b2591 & 5 \\
84 & fa87fce54c5e3d9964a49 & 5 \\
\showrowcolors
\bottomrule
\end{tabular}
\end{center}
\end{table}

\begin{table}
\begin{center}
\caption{Reached near-optimal solutions}
\label{tab:ReachedNearOptimalSolutionsPartII}
\ttfamily
\rowcolors{2}{light-gray}{light-cyan}
\begin{tabular}{lll}
$n$ & Sequence in HEX & PSL \\
\toprule
85 & 1007b4a2f86ae1cc4cb36 & 5 \\
86 & 1378ae9166656250f0435f & 5 \\
87 & 1850253c557b83626f3369 & 5 \\
88 & 2c43c8691299154d4fbf04 & 5 \\
89 & 17b237ec7daea0c1a7d8d4e & 5 \\
90 & 7ca2b17db11f675bf5ad30 & 5 \\
91 & 3dffa15d0b98b16c5c65349 & 5 \\
92 & 4c9254743cf393b942217f & 5$^{\hourglass}$ \\
93 & 1c9cdc87fdfa50e348aab25c & 5$^{\hourglass}$ \\
94 & 3c144be5d296b3e65dc46600 & 5$^{\hourglass}$ \\
95 & 18d0d61707462bd427fedb24 & 5$^{\hourglass}$ \\
96 & d02532058d9cf0d019e578aa & 5$^{\hourglass}$ \\
97 & 1a542ff2c2ee6feb3b065186c & 5$^{\hourglass\hourglass}$ \\
98 & 1e61c02e1104b6ea5a981cdc9 & 5$^{\hourglass\hourglass}$ \\
99 & 71cf7d0426a0646b20d8a972 & 5$^{\hourglass\hourglass}$ \\
100 & 191f7308a8fc4fac34a902c90 & 5$^{\hourglass\hourglass}$ \\
101 & 1ff41f8ee334912c6cdca8d28a & 5$^{\hourglass\hourglass}$ \\
102 & 40477758e393668697c0fd6ad & 5$^{\hourglass\hourglass}$ \\
100 & 255b559207e991908213c63e3 & 5$^{\hourglass\hourglass}$ \\
101 & 1ff41f8ee334912c6cdca8d28a & 5$^{\hourglass\hourglass}$ \\
102 & 40477758e393668697c0fd6ad & 5$^{\hourglass\hourglass}$ \\
103 & 4fbcf31f8fe6d103ea8dacad48 & 5$^{\hourglass\hourglass}$ \\
104 & 2e76361a08417ada07987744dd & 5$^{\hourglass\hourglass\hourglass}$ \\
105 & 199bb906d3e822bc96a4110e1c7 & 5$^{\hourglass\hourglass\hourglass}$ \\
\showrowcolors
\bottomrule
\end{tabular}
\end{center}
\end{table}

\subsection{Finding binary sequences with record-breaking PSL values heuristically}

The results achieved throughout the experiments described in Sections \ref{subsec:1} and \ref{subsec:2} demonstrated the  efficiency of Algorithm 2. Thus, we have further launched the algorithm on binary sequences with lengths up to 300. The results are given in Tables \ref{tab:106-300partI}-\ref{tab:106-300partVI}. The binary sequences with record-breaking PSL values are further highlighted with the symbol $\blacktriangledown$ (black triangle pointing down). Almost all of the results known in the literature were improved. More precisely, we have improved 179 out of 195 cases. Curiously, for some lengths, we have even revealed binary sequences with record-breaking PSL values, having a distance 2 to the previously known PSL record value. We will mark those improvements with a double black triangle symbol. An example of such length is 229.

In \cite{coxson2014adiabatic}, the best results achieved by the D-Wave 2 quantum computer for binary sequences with length 128 is PSL 8, while Algorithm 2 could reach PSL 6 (see Table \ref{tab:106-300partI}). For longer lengths, for example binary sequences with lengths 256, the best PSL achieved by the D-Wave 2 quantum computer was 12, while during our experiments we reached PSL values of 10. In fact, we reached PSL values of 10 for binary sequences up to 271 (see Table \ref{tab:106-300partIV}). For completeness, since the D-Wave 2 quantum computer is tested on binary sequences with length 426, we have further launched Algorithm 2 on the same length. Surprisingly, the algorithm was able to find binary sequences with PSL values of 17 (the best value achieved by the quantum computer) for less than a second. In fact, it reached PSL values of 16, and even 15, for less than a second as well. However, PSL value of 14 (see Table \ref{tab:longerSequences}) was noticeable harder to reach (199 seconds). During this optimization routine, and driven by the results provided in Table \ref{tab:alfa500} (since 500 is close to 426), we have updated the $\alpha$ value to 3. 

Recently, in \cite{coxson2020long} a multi-thread evolutionary search algorithm was proposed. By using Algorithm 2 we were able to improve almost all of the best PSL values from the aforementioned paper - usually for less than a second. For example, the best PSL value for binary sequences with length 3000 achieved in \cite{coxson2020long} is 51. We have launched Algorithm 2 on binary sequences with the exact same length. It should be emphasized (see Tables \ref{tab:alfa2048} and \ref{tab:alfa4096}), that the $\alpha$ parameter should be increased to 6. Record-breaking PSL values of 44 and 43 were reached for respectively 111 and 371 seconds. In Table \ref{tab:longerSequences} an example of such binary sequence (2nd row) is given. The last column of the table provides a more quantitative measure of the record: $\blacktriangledown$x denotes that the corresponding binary sequences posses a record-breaking PSL equal to $P-x$, where $P$ was the previously known record.

\begin{table}
\begin{center}
\caption{Binary sequences with near-optimal PSL - part I}
\label{tab:106-300partI}
\ttfamily
\rowcolors{2}{light-gray}{light-cyan}
\begin{tabular}{lp{6cm}l}
$n$ & Sequence in HEX & PSL \\
\toprule
106 & 35101a2373a0160d982f6b4e39a & 6 \\
107 & 2408504b2beac46b8d93cc85f86 & 6 \\
108 & 727184e79679234058155e880bd & 6 \\
109 & 5db00f58363f65c08452544632b & 6 \\
110 & 2b5085f188c82cbb79e1ae25c1bb & 6 \\
111 & 700f7ceb4b8a926c793caafcdcee & 6 \\
112 & 1c62bf5e0e2bf9bdb9db524d921b & 6 \\
113 & 10e8e632f9a52d803cd7eac6eddd5 & 6 \\
114 & 3fad9a9fa616431ee6a6b8746ba74 & 6 \\
115 & 637c6cdec32bd4cbaecaf2ffe1610 & 6$\blacktriangledown$ \\
116 & a03feff259d626e9c4f46471a5168 & 6$\blacktriangledown$ \\
117 & 1b33da4cc6d5dc7f8a55c9007cb8f0 & 6$\blacktriangledown$ \\
118 & 23c598f4ac7f6afde47b84c05dd592 & 6$\blacktriangledown$ \\
119 & 60835d6bb25f775d6b588d9e361f81 & 6$\blacktriangledown$ \\
120 & 98cc2e429c2f810668dfdf14bab0b2 & 6$\blacktriangledown$ \\
121 & 178ffe7181c3f443365313724aac95a & 6$\blacktriangledown$ \\
122 & 30d4e9ae516cf0320ad003177377485 & 6$\blacktriangledown$ \\
123 & 369ec917afe507e53bdc97151138738 & 6$\blacktriangledown$ \\
124 & f15ce151edfd7f0ca9eb4496d833233 & 6$\blacktriangledown$ \\
125 & 1b8730333bcdf414d92203c581a554a5 & 6$\blacktriangledown$ \\
126 & 3b9275a7ba7661bb8dbf8e078ad41257 & 6$\blacktriangledown$ \\
127 & 2933b32d40937c4b6f08e03a851c2c2a & 6$\blacktriangledown$ \\
128 & 84528942da6f07e733404ee8ba70c3ae & 6$\blacktriangledown$ \\
129 & 1f80f99bf3cc5c3d6f1aacd4209aa925b & 6$\blacktriangledown$ \\
130 & 2678ae07e71929fb587022ed6bfdb576d & 6$\blacktriangledown$ \\
131 & 3cbf4b091ea86cea277167ac6304c812 & 6$\blacktriangledown$ \\
132 & 410028af0ea52e93f029f908ce74d8c99 & 6$\blacktriangledown$ \\
133 & 10c27978f1888d4fb0a97c9326ecafe97f & 6$\blacktriangledown$ \\
134 & 3f01b89e464dccaabce38e920492b56810 & 6$\blacktriangledown\blacktriangledown$ \\
135 & 550c944868887c4b7b8709d8263de6c81a & 7$\blacktriangledown$ \\
136 & dc789e3aa4f65db16085033ab4b40aee42 & 7$\blacktriangledown$ \\
137 & 1bdfe2817aaa3b39d39daf366d86bc0f49 2 & 7$\blacktriangledown$ \\
138 & 1e618e9ba6c707dc94f05ad723357b2bff d & 7$\blacktriangledown$ \\
139 & 2bd70f3cde89ad5316439120fe3b9b480b 5 & 7$\blacktriangledown$ \\
140 & 79f8036d08785fbef98ba3b2eb54652eb3 3 & 7$\blacktriangledown$ \\
141 & fdf5f77808f6cf055b0dd9295c878ad32e 2 & 7$\blacktriangledown$ \\
142 & 343638ce5ed915e8abcc9a0beef8128148 94 & 7$\blacktriangledown$ \\
143 & 554e194c63ca5a65f47de2fd999fc0227e bd & 7$\blacktriangledown$ \\
144 & e757f83fd667a6c479d5296908879f6c8d 2e & 7$\blacktriangledown$ \\
145 & 1051d14d00c893e49498fdba0570862f53 9ca & 7$\blacktriangledown$ \\
146 & 1c8f3f584efe71220e0da5d4d58d10ed11 ec1 & 7$\blacktriangledown$ \\
147 & 2072a669eade89e6058251c9cc2628a5f5 602 & 7$\blacktriangledown$ \\
148 & 136cbb11363a7078d55f1dc696f217b588 520 & 7$\blacktriangledown$ \\
149 & 152214204e428bf4553661919fe41c0db6 9e18 & 7$\blacktriangledown$ \\
150 & 4fc361d2f104c9510a54c53e9afa612346 7f6 & 7$\blacktriangledown$ \\
151 & 3b93bb695ed592557b82497047438f87c4 318 0 & 7$\blacktriangledown$ \\
152 & 62913b08d6326c46c082e52e3feb0e0b6d 7505 & 7$\blacktriangledown$ \\
153 & 152b80f95df5a4a0e1a2e30cbf68cf76e9 ccdeb & 7$\blacktriangledown$ \\
154 & 183f0383fe80cccbf38ac495965dae7bd7 9a695 & 7$\blacktriangledown$ \\
155 & 1431f0be9440e92cdd1c4d5680659df937 7adca & 7$\blacktriangledown$ \\
156 & bbbd712a19673174fdbc9ad3e78d06f40b b1a84 & 7$\blacktriangledown$ \\
157 & 1444313cfc12546b26e36eb70568dc11b7 06bcf5 & 7$\blacktriangledown$ \\
158 & 18785b52d7074935b31708ef988f769911 040aa7 & 7$\blacktriangledown$ \\

\showrowcolors
\bottomrule
\end{tabular}
\end{center}
\end{table}

\begin{table}
\begin{center}
\caption{Binary sequences with near-optimal PSL - part II}
\label{tab:106-300partII}
\ttfamily
\rowcolors{2}{light-gray}{light-cyan}
\begin{tabular}{lp{6cm}l}
$n$ & Sequence in HEX & PSL \\
\toprule
159 & 41c5d5f8d8012c40f00d6ba24d35a539cd 8a573a & 7$\blacktriangledown$ \\
160 & 7277c5d1140ae6e5638c47ab40937830f2 1b684b & 7$\blacktriangledown$ \\
161 & 1c9d7e8ec413f2eddacc7be1a45a4318ba 3ab2b46 & 7$\blacktriangledown$ \\
162 & 36280d42653b385e990c70aec3d64845dd 59413e & 7$\blacktriangledown$ \\
163 & ff4a2a50fadfb069cb64a79bb8eafdf55e 660cc4 & 7$\blacktriangledown$ \\
164 & 8cbd592237b8e9d5dd7fddb148e13a7c0e f03696c & 8 \\
165 & bdfe78f96cd0a73e41fa8764667b9e82d1 54d6544 & 8 \\
166 & 12c242012b761f803271ab9649f67432ee 288d398a & 8 \\
167 & 349aab4a5752c6459e4cd43f18708fd980 18044fc3 & 8 \\
168 & a18f9ca18bdb8eaf44a84db7f3f92ddd36 0ec23bc4 & 8 \\
169 & f77f9c30338bec86cb76455ec4af4d4394 769e17de & 8$\blacktriangledown$ \\
170 & 1647c513e17c8b5ac12f169cf4008e77de aeedd71d9 & 8$\blacktriangledown$ \\
171 & 455bce3cd34aa5199a53b3f9900ed684d8 11607828e & 8$\blacktriangledown$ \\
172 & 8a862f714aa517b5e1d4c9e784b66d0c07 eccfd9f61 & 8$\blacktriangledown$ \\
173 & 9087c81b16785b95b1f63942ab8829d1da e83048267 & 8$\blacktriangledown$ \\
174 & 3a595abedb5fb13e998250683feaa608f1 b10f721e8c & 8$\blacktriangledown$ \\
175 & 75f9a9db5111b640009d36bc18a71887a8 d4f60f5079 & 8$\blacktriangledown$ \\
176 & f8b81eb83c80faa526c53d6c43bbb18d34 c2b7df7bb3 & 8$\blacktriangledown$ \\
177 & ceca7a7c3d3d4ed8893081464daa5d50cf 40905fe630 & 8$\blacktriangledown$ \\
178 & 240603787909825762b567fe0a338e0aeb 85db46e98a4 & 8$\blacktriangledown$ \\
179 & d27d4a3d8ce26560a137f967fd5b2a22fe 7ea4e9cb1e & 8$\blacktriangledown$ \\
180 & 45f880bba360f4fe67321649f77be67971 e729d54b4a5 & 8$\blacktriangledown$ \\
181 & 16e90d659806f9ad47ab399481f7975581 4e0cc9805074 & 8$\blacktriangledown$ \\
182 & 2ad575a76ca6eb9f36b207790cec2047db f70dd1f07095 & 8$\blacktriangledown$ \\
183 & 3080a6b5d518e2437cbe03e83276091f3d 9ad5bb717275 & 8$\blacktriangledown$ \\
184 & bd09e4073d719c5290dc81d3edc090b050 3345a2aaddb7 & 8$\blacktriangledown$ \\
185 & 1b0245da96f15f3faaf6f0e71c5a6c6e2e 7ba4fa2190aff & 8$\blacktriangledown$ \\
186 & d6c3079d747a0496d4eab7337a91236c73 cefba0eff4f4 & 8$\blacktriangledown$ \\
187 & 34999dc9c93025871f7aaceb517d0e451c 07b504a75da01 & 8$\blacktriangledown$ \\
188 & 36766a797988a55100a42a91e73c43f005 b76d60705f364 & 8$\blacktriangledown$ \\
189 & 880723487ff2acbc3e65d1eba13327b9a0 5965bd52d14e7 & 8$\blacktriangledown$ \\
190 & 21e50af105ba1d87a44214221d935bba27 35951f776101cf & 8$\blacktriangledown$ \\
191 & 6122466d46065abb2e2595ed350f45d4a7 f173881f4c33ef & 8$\blacktriangledown$ \\
192 & 1fbfab7bc285711fb852eb5f00b2ba9c36 98e27cd26a66c9 & 8$\blacktriangledown$ \\
193 & 11e5e2e1ea52cd9c13f6ec031979a99549 b90fb8c2600a288 & 8$\blacktriangledown$ \\
194 & 35d745068d86b74ca0a6d8c73a39676ea7 7bd2b4bc0fc0267 & 8$\blacktriangledown$ \\
195 & 1c841bd699c259b0d801b20e4fd8bebe1c 6567ae3abd08a95 & 8$\blacktriangledown$ \\
\showrowcolors
\bottomrule
\end{tabular}
\end{center}
\end{table}

\begin{table}
\begin{center}
\caption{Binary sequences with near-optimal PSL - part III}
\label{tab:106-300partIII}
\ttfamily
\rowcolors{2}{light-gray}{light-cyan}
\begin{tabular}{lp{6cm}l}
$n$ & Sequence in HEX & PSL \\
\toprule
196 & b6a64ce8063c6116f91dd3cfc332f8aac5 f7bdc8a0bad6d2a & 9$\blacktriangledown$ \\
197 & 2b7ceef5fba16ec29257b30a65a26ac34f 1841ddc7c0e0de7 & 9$\blacktriangledown$ \\
198 & 2da669214cb962a811544e5d3d37a000f8 c0c60dced1ed0bce & 9$\blacktriangledown$ \\
199 & 1144b275da9c8adb8fffce37c87ba0d2c3 c6bda983f4dc032b & 9$\blacktriangledown$ \\
200 & 66c30c122f4ee5d8b01ab9155a1ca5afed 0d37d4df0775bd84 & 9$\blacktriangledown$ \\
201 & 82a1c892ca09589a5f1ba194c682ef0f71 d182378a64895ff4 & 9$\blacktriangledown$ \\
202 & 1d045e3d7d3e006c938fb456d5f2a4bf5e 4dcce9c41ca663186 & 9$\blacktriangledown$ \\
203 & 6413fc8964522104171ca948e5d4c4e1cf ade1a82d03b3e640d & 9$\blacktriangledown$ \\
204 & 6730c61d894ad6db47d7db1707d109a8fd 7e9912cfee2df887d & 9$\blacktriangledown$ \\
205 & d24ff6dfd7766450d28c6f1d08aa13c6f5 060b93ef182d5e847 & 9$\blacktriangledown$ \\
206 & 7372ccbe4d517dc500e9ed586a99c9fc60 a442016a06fd0c961 & 9$\blacktriangledown$ \\
207 & 5b92dad3371cc960e08e1993a80ac0a9f5 73c2708165ba02bf5f & 9$\blacktriangledown$ \\
208 & d47fffd42e8257a630ef1673359f05eb26 ce173462e0ecb498d2 & 9$\blacktriangledown$ \\
209 & 1bf64d73ea8531230afd6c614fceee5aad 2714cd7c1674125f01e & 9$\blacktriangledown$ \\
210 & 31da42975a5a3c741f6506fc77598874bf 77e37f2ae29fb1304dd & 9$\blacktriangledown$ \\
211 & 252e50a7cd40fd82e13aae3096361608b2 3030076fbbd84ca636e & 9$\blacktriangledown$ \\
212 & 87d63ff093d2c932221b74ae6e9443ac63 33b42e0b890a5754141 & 9$\blacktriangledown$ \\
213 & 54614010e30c87b5366b6baa6400fc7e8b 57067a894e9b3f898e4 & 9$\blacktriangledown$ \\
214 & e7d4a6d69fda9cf9843db94242a88c5cd7 77cd24165c2f913e0fc & 9$\blacktriangledown$ \\
215 & 301c7898c56aa56687800ffbf3e65a5787 6867c9426eb3dd5d46d3 & 9$\blacktriangledown$ \\
216 & d332ccdcab19af1972f93007baf8af8057 c3af4b59e4d040624b52 & 9$\blacktriangledown$ \\
217 & a87867118f48a6922d161093f015d7f8dd b57c80cb5aedd1b0b177 & 9$\blacktriangledown$ \\
218 & 4c91d36554864c73c5ae223a17dd60ec62 96849685d7fb81f3f881 & 9$\blacktriangledown$ \\
219 & 536a2df324baa32c8488880d9ae152f5bd 0b808ebcf131fc0c293c7 & 9$\blacktriangledown$ \\
220 & bc39257be78b79101abf2c3edb9b3c01e4 157240d46a6319c5789d2 & 9$\blacktriangledown$ \\
221 & fadda9f6d109fcc882a91bab8478e6ebc5 713826d19fd06b485a061 & 9$\blacktriangledown$ \\
222 & 1e1a28caa7a070d16e6300965eba9752b3 4e37e66d02139025cfc84f & 9$\blacktriangledown$ \\
223 & 49e28e14ca6daa3c6fc973368464a08c55 94bf408129cfa607b303d1 & 9$\blacktriangledown$ \\
224 & b5a435ab97a31e722120bbf812cd251cc7 032281cc0aa29f07f9b66e & 9$\blacktriangledown$ \\
225 & 1f61bc4168c021782f9e50c6b52a1c546a da0864fd9313b32a9bfae98 & 9$\blacktriangledown$ \\
226 & 3bbbf1c19f6ec551ff0ab982ab334dae01 29a63cc6b58b61e968d0d80 & 9$\blacktriangledown$ \\
227 & 378320814f8021439fe15e5a12add18b76 0cd0788aba8ed3630926333 & 9$\blacktriangledown$ \\
228 & ac5471683c569456b21141ec539fa32e00 78998f3800d377665b36fa & 9$\blacktriangledown$ \\
229 & a18ea64e0c7d887c6fb51278b686a8b401 66199ef8050906a7c1516b3 & 9$\blacktriangledown\blacktriangledown$ \\
230 & 25474a4ba6e7c3434d1c724ef643ea3181 2c8fa1bbfc877bf6ee5488f7 & 9$\blacktriangledown\blacktriangledown$ \\
231 & 180e1d36289672c3086a1511d58dfeb4a7 f13ca44b44fec5d024664dde & 9$\blacktriangledown\blacktriangledown$ \\
\showrowcolors
\bottomrule
\end{tabular}
\end{center}
\end{table}

\begin{table}
\begin{center}
\caption{Binary sequences with near-optimal PSL - part IV}
\label{tab:106-300partIV}
\ttfamily
\rowcolors{2}{light-gray}{light-cyan}
\begin{tabular}{lp{6cm}l}
$n$ & Sequence in HEX & PSL \\
\toprule
232 & 886bf85a5bf40b7b2fab51ee8712e2bba7 5b358384435d3ccc99dbf7e6 & 9$\blacktriangledown\blacktriangledown$ \\
233 & 5dda7518a3629e66f6ec3823f6cc6c373b 4bac795efacf416fe1a1ab00 & 9$\blacktriangledown\blacktriangledown$ \\
234 & f8e141d03a5bb1d91ce20721cdb6207f56 b699d33bf575955694dfa930 & 10$\blacktriangledown$ \\
235 & 3feaa21651ef22c8cb05ab35df33b138a0 8c83e1a1ed24685592035e152 & 10$\blacktriangledown$ \\
236 & 5453ea9ffc1e60c3285de3d07b64a1bcc0 95366d4c437dfccd5d8f4fcea & 10$\blacktriangledown$ \\
237 & 165567767124fabcb4d08f0da7140abd81 f42e5c9a831dda76fffd894c71 & 10$\blacktriangledown$ \\
238 & c81ea65bf4b9df2e7f7066454c2d3c8e6a 2841e27963c8229db40a0afd8 & 10$\blacktriangledown$ \\
239 & 6a66b95e25a3cb20e16c7b36b1b22e5988 21242ffc69eeaed03bf9f9d753 & 10$\blacktriangledown$ \\
240 & dcd3bec7a1856d4ea4febb5c0dcc52e119 ffaa69d4c86df1470530793374 & 10$\blacktriangledown$ \\
241 & 15cabbcb3c965d13d1baf6581833a05593 c6ff73c18dfca9e96272a467f29 & 10$\blacktriangledown$ \\
242 & 2cf51c98f2793804326afb59471b2243a9 12fa50b7abce08ef22607d03941 & 10$\blacktriangledown$ \\
243 & 66346d9c9f2d3393fdeaa0075e7f573ef2 7a1b4b8630b4a322df02f9ba47d & 10$\blacktriangledown$ \\
244 & cde4bae1750d2d31e5e3b193df44580f92 245b262ffaaf6e42c6bde7b9532 & 10$\blacktriangledown$ \\
245 & 1af4e7a8ed850811188970ae2af8180736 3afb0113d0f9166b49916df928d6 & 10$\blacktriangledown$ \\
246 & 30b36a460dbd8ab690c173b8d8ca8c0351 cb3a170bba020a9417843dd76dd3 & 10$\blacktriangledown$ \\
247 & 5de6adef6aa7775d3812b0cd7831689b5e 39682e61899e9ba3f039b00e27b4 & 10$\blacktriangledown$ \\
248 & f41f437cb07cf0a0aadf0c67b3f7fe114c c66b766ccd153185293943089549 & 10$\blacktriangledown$ \\
249 & 10242f665effd3eb4875a1fab42f9d4515 fc9e251dae3c607319a69e49366ef & 10$\blacktriangledown$ \\
250 & 2007616f89095843f3ced5634bf501cfff 55adb4589658662e8ba374f65c676 & 10$\blacktriangledown$ \\
251 & 275419d5069976e3bdde14b3329284641e 6164276b8012963f4d383e161fcba & 10$\blacktriangledown$ \\
252 & 1b55a5dadcac2ee8c3ef41026edcc98eab f592878208e314f6349886407e13d & 10$\blacktriangledown$ \\
253 & 16fa06a49b5776c2a804a3f64b59e4fd20 3a358e8a77d8f79f159d7c34654e60 & 10$\blacktriangledown$ \\
254 & 20d5c99925b7a51f543e49ff428d5d4e54 8a26e1280a1a2d9fc5cc33018c70ce & 10$\blacktriangledown$ \\
255 & 10008133c4e8b9aa47e1546b8b75a0a4fb cc1d2c7925637235e4866f23d20cf2 & 10$\blacktriangledown$ \\
256 & 6e6053b51d9f80a561e97e2cc13cae1d56 38728f2013377e867fbbee26bada65 & 10$\blacktriangledown$ \\
257 & 1a24b6e6c465cf993425fe01cb10c2ac88 2285c51cea5697d378bc40305c6e753 & 10$\blacktriangledown$ \\
258 & dafc4a13dbc909c653b76970b24085986f f0fd93e73d6bcd8bba9aae855ce8af & 10$\blacktriangledown$ \\
259 & 23c7a45f27e3ff8fb66d31f630620d9f6f 959318ea2754cff5256657508bdad2c & 10$\blacktriangledown$ \\
260 & 94db24992764caf16520a31303c3d0a967 2e74e01e8012d787381aaaeae319def & 10$\blacktriangledown$ \\
261 & 28d24a7097956e9f7a63b183c0d97211ee 4f99b9f94a3de360ff75f75082fb55d & 10$\blacktriangledown$ \\
262 & 2547150b862f86ac277033a8d7de1cfd81 8cd1012db7104817cbf15c29924695c9 & 10$\blacktriangledown$ \\
263 & 58333ccc921a4318cbddf299f4a0d055a3 3a13554056c856a9380b4ff0e1c60d3f & 10$\blacktriangledown$ \\
264 & 7a4dd00f8bbafc5095a2f5f00da7131ba7 d6f7ac4ce20662e388a6b0c21273204b & 10$\blacktriangledown$ \\
265 & 154ab3ecf9568391efd8918b059f988d67 a21805a46107cb6b89bd30f4c47405c51 & 10$\blacktriangledown$ \\
266 & 3c690152ba0daf7d5b4f7a3ee3c88ab33f 6bb8252dc786c8ccd668169c4bbc4cfc2 & 10$\blacktriangledown$ \\
267 & 15e1810bfa1308e523b851c7078b2be464 f66df69c7492775594b91644a16e77aff & 10$\blacktriangledown$ \\
268 & 32c38e387faae3e8b74eb7d4675bfa49f5 00cac6c56b4de44a8b7f9d8372666090b & 10$\blacktriangledown$ \\
\showrowcolors
\bottomrule
\end{tabular}
\end{center}
\end{table}

\begin{table}
\begin{center}
\caption{Binary sequences with near-optimal PSL - part V}
\label{tab:106-300partV}
\ttfamily
\rowcolors{2}{light-gray}{light-cyan}
\begin{tabular}{lp{6cm}l}
$n$ & Sequence in HEX & PSL \\
\toprule
269 & e6fbaa465ee2a294646b484fbf7d498512 837f32dc48de2872f0781741941892680 & 10$\blacktriangledown$ \\
270 & ca6d5d2e2898349ca6f36814244ad6e204 8fc50210d8a0fe07fcd5d7f135261c718 & 10$\blacktriangledown$ \\
271 & 4fd6ffb4ef673619e25b08bdb8157332e6 15587d1c72ee2d9302c5feb706b0acdab4 & 10$\blacktriangledown$ \\
272 & eb8a2f2227f60eca7d47a60d44193beef1 4b2502f3b5a198f69d3ed7dfb4eca72b41 & 11 \\
273 & 1b007f99648f37cf3f43ffdb61260d2d33 b65231ad1cbc3353a1ec6e4bc5555d5ab9 5 & 11 \\
273 & 1d92f5d3696863c9fa0972f85e9023bf72 63e0d0472f3a817d42462388332a9db3ba d & 11$\blacktriangledown$ \\
274 & 2a377a8cd8e836fc187135c97cb4f69fad ef3a4367b96014b1b9a79bb40d1120baa7 5 & 11$\blacktriangledown$ \\
275 & 5991082785400a4fec7053b34aeba361d9 542b51c7533d37b28524c29f747f285b8c & 11$\blacktriangledown$ \\
276 & af8eb78a4018df61ad9e2d5c980dd38ea4 dbd3cc1d37126245796adbfad9dccfe27 & 11$\blacktriangledown$ \\
277 & 4a97467cb36e66d3c4062908017d0aa39c 6a04ad0f2f27b6c10b1dbaec226a396dc1 d & 11$\blacktriangledown$ \\
278 & 96655611a994569ea5924430f8fbaace17 8f1df22f07a48c180bef02336e65223642 b & 11$\blacktriangledown$ \\
279 & 7425c1ec9da091b4d0ee98297cf8a600cb b43c455e0031c4f15c642251892bdbcbd5 d7 & 11$\blacktriangledown$ \\
280 & 55bef3e1c6a79c1a03aad609724c2da00b ba2ad6484112fe95db18d81f99948c6f0b 32 & 11$\blacktriangledown$ \\
281 & 1424e102f4fde1aa05941514283b49ba3e 1786dae904facc4c6db8ac6632ef12acc8 9ce & 11$\blacktriangledown$ \\
282 & 3e88983536cd657fd8069a3360e796e35a dc35cb8ab5c6f0ebeeefee25ad9daecc06 81f & 11$\blacktriangledown$ \\
283 & 5f08ad54acf01756103661f0e35a1c815c d9465bf0909bdcbca2081b5ce79bbb96e7 4d & 11$\blacktriangledown$ \\
284 & d1195f2440f108379fb13357dd894f83ff 89e0e313269f6b5f48f675acb1218d5769 aeb & 11$\blacktriangledown$ \\
285 & 1fd2b9a522fa0ba0363cefa32874704a1a f558b374eb3eadff9593c7925bbc98e4f6 62d7 & 11$\blacktriangledown$ \\
286 & 12410ee79818cd60c7230ad510aadec039 2e476e0f0a036f167bf2be2cab02c0d6d4 4d81 & 11$\blacktriangledown$ \\
287 & 63dac2f781e251694b5e9978aa03ca24f2 a20ad51bbba930e99d93590608f330099d e606 & 11$\blacktriangledown$ \\
288 & 69a4d15a39cd274d62e3f41c235b3b280f 0336af0833a646b21eb0a04085c40b5fab 1aae & 11$\blacktriangledown$ \\
289 & 16d9909ffc2421af02de219e1d86e042cd bfaff9be97237531d2e1ab96739b5eab8b 166aa & 11$\blacktriangledown$ \\
290 & 15c7ff0aef22f9dbdf7394c8094b13871a c35a9bcd81a472251e5024efd3605951fa 0d157 & 11$\blacktriangledown$ \\
291 & a1c202c94a731846da997686016197dbcd 6a6ca7cc264437646ac0c0fcbd11ff5ae0 7555 & 11$\blacktriangledown$ \\
292 & 6e7871e089cc8db9274cf3a22f22d3d452 3d272db2952ab2ac27188b6fbf47faef01 bbdf6 & 11$\blacktriangledown$ \\
\showrowcolors
\bottomrule
\end{tabular}
\end{center}
\end{table}

\begin{table}
\begin{center}
\caption{Binary sequences with near-optimal PSL - part V}
\label{tab:106-300partVI}
\ttfamily
\rowcolors{2}{light-gray}{light-cyan}
\begin{tabular}{lp{6cm}l}
$n$ & Sequence in HEX & PSL \\
\toprule
293 & 4e6aa8af6a0ead457af0ad0ca55efe940e 310e4e6f21b73cf0006124c90360db0b98 7db6c & 11$\blacktriangledown$ \\
294 & 1a5b6d16eca315188a6c5271c7a7ab9eb3 a5ee0efdaabf07b9578110e7fcfe06ccd0 a47ecc & 11$\blacktriangledown$ \\
295 & 4707bfcc051613d674df4982da568161be 90f8cb12bf339535d1a7488f0468c03112 ae1157 & 11$\blacktriangledown$ \\
296 & b71a2ac9b154ef459223e3b03cd2394c7e 3c15f1ac6a2272deea2c235a20fb4bdbdf 3b7fb4 & 11$\blacktriangledown$ \\
297 & 2ff21574c741f68aa9d0872b97acc757c6 fdb374791dee45c4a41c274d6c6df59200 62de92 & 11$\blacktriangledown$ \\
298 & 32c7b45b87be1884edfec5712d7e3efcef 2e825460a6d5dc1b9d4335581e4e33b454 d9126b4 & 11$\blacktriangledown$ \\
299 & 587814353b51d6b1d029f7fe73bd9c88ad 984a394357ee609923a3ec1923e0bb4047 492ea83 & 11$\blacktriangledown$ \\
300 & 5b2a550cad8a468cac4ac82be6ad333849 c865361c6d800818ef9387d6513e8cb81d 0f23ff6 & 11$\blacktriangledown$ \\
\showrowcolors
\bottomrule
\end{tabular}
\end{center}
\end{table}

\begin{table}
\begin{center}
\caption{Example of long binary sequences with significantly better PSL values}
\label{tab:longerSequences}
\ttfamily
\rowcolors{2}{light-gray}{light-cyan}
\begin{tabular}{lp{5cm}ll}
$n$ & Sequence in HEX & PSL & $\blacktriangledown$  \\
\toprule
426 & 3075e0e3e3c1581d2af808dfee48904 226a942d671d897292c4613c5b19a5d d22a6799309414418db4ba724a9fd8f fd1dd109b71493 & 14 & $\blacktriangledown$3\\
3000 & 9c9d1dd018fecf19c744616ad4b166 50e04945bf3f38486f3e52499f8687b d6a090f4b79735ec64f9987f6ac4985 ba941983ecdb9d1d0fd861dfdc7ed4a dd34ac12f08b559aa8c22cf70b4724a d819dbb4ddf4678582db3601786cc56 7f8d290b90cbf46e2939152989bba06 5e1644ec8b1d995e9d8d68221ff5166 66bff43b0a993eaa9ba440f0b79f00e 083acd93b1b64ee5acc52cb1a3bc77e 7c01a14a8a7d8003c62fc5778be6a05 df09b9fc03b70dc2df6850a61ed7045 398c52aa1b5baf036848553d7dd27f8 cb72ed847c6796f7216a975dc497149 ef6eab576508ac77dc3c8837d54d952 1d151694dea17e2bb4969a2c4461616 fafaacb172e35685b3bd63152287a79 e329c65b01a41030bf595ec7ef87188 b37a4d3552e73fadefcdf57b05cc618 904a2fdfd52ff7e8a8c1ea9fdf9db08 957495f01fd6ca7ff219ae3c4624100 d4eee30cc0db5aa8e9f548c31b10593 f138b2c7d22c3f7c16279b7b2f65de7 d17494944967d341c6c0c4e70863b00 201984a & 43 & $\blacktriangledown$8\\

\showrowcolors
\bottomrule
\end{tabular}
\end{center}
\end{table}

\section{Hybrid approaches for PSL-optimizing problem for long binary sequences}
\label{sec:hybrid}

The reasoning behind announcing one binary sequence as long, or short, is ambiguous. Measuring the largeness of a given binary sequence is probably more related with the capabilities of the used algorithm than the actual length itself. From practical point of view, some algorithms, or their implementations, would not even start the optimization (or construction) process, since their computational capabilities (or hardware restrictions) would not be able to process the desired length. For example, as discussed in \cite{coxson2014adiabatic}, the usage of a 512-qubit D-Wave 2 quantum computer limits the code length that can be handled, to at most 426, due to a combination of overhead operations and qubits unavailability. Moreover, it was estimated that a 2048-qubit D-Wave computer could handle binary sequences with lengths up to 2000. Hence, the exact fixed value differentiating short from long binary sequences is still unclear.

In Table \ref{tab:mSeqComparisonTimes} some detailed time measurements of binary sequences with lengths $2^g-1$, for $g \in N, g \in [13,17]$ are given. The binary sequences are specially chosen to exactly match the lengths of the well known m-sequences, generated by some primitive polynomial of degree $g$ over $GF(2)$ denoted by $\mathbb{M}$ (see \cite{dmitriev2007bounds}) and the  binary sequences generated by Algorithm 2 denoted by $\mathbb{A}$. The $\alpha$ parameter was fixed to 4. The last column ($\mathbb{A}$) denotes the time needed for Algorithm 2 to reach the corresponding PSL (s, m, h and D denote respectively seconds, minutes, hours and days). Evidently, the longer the m-sequence, the harder for Algorithm 2 to find a binary sequences with better PSL value is. For example, Algorithm 2 required approximately 3 days to find a binary sequence of length 131071 with lower PSL than the optimal m-sequence having the same size. Given a PSL-optimizing algorithm $\mathscr{A}$ we will reference the length $n$ of a binary sequence as $\mathscr{A}$-long if the expected time from $\mathscr{A}$, starting from pseudo-randomly generated binary sequence with length $n$, to reach a binary sequences with PSL $p$, s.t. $p \leq \floor{\sqrt{n}}$, and by using single general purpose processor, is more than 1 day. Otherwise, we will reference it as $\mathscr{A}$-short. Throughout the radar literature statements that the asymptotic PSL of m-sequences grows no faster than order $\sqrt{n}$ were frequently made. However, as shown in \cite{jedwab2006peak}, this assumption was not supported by theory or by data. Nevertheless, it appears that the PSL-optimal m-sequences are very close to $\sqrt{n}$ (see \cite{dimitrov2020aperiodic}). Thus, the threshold value of $\floor{\sqrt{n}}$ is based on the expectation that the optimal PSL value for a given binary sequences with length $n$ is less than $\ceil{\sqrt{n}}$.

From now on, we denote Algorithm 2 as $\mathscr{A}$ with fixed $\alpha$ value to 4 if not specified otherwise. During our experiments and by using $\mathscr{A}$, we have reached to the conclusion that all binary sequences with lengths $n$, s.t. $n > 10^5$ are $\mathscr{A}$-long. In this section, we have investigated some hybrid constructions which could be applied in those cases when the binary sequences are $\mathscr{A}$-long. 

\begin{table}
\begin{center}
\caption{Time required to find better PSL values compared to known results from m-sequences exhaustive search}
\label{tab:mSeqComparisonTimes}
\ttfamily
\rowcolors{2}{light-gray}{light-cyan}
\begin{tabular}{lcccr}
$g$ & $n=2^g-1$ &  $\mathbb{M}^\mathbb{F}_n$(PSL) & $\mathbb{A}$ (PSL) & $\mathbb{T}$ \\
\toprule
13 & 8191 & 85 & 84 & 19s \\
13 & 8191 & 85 & 83 & 23s \\
13 & 8191 & 85 & 82 & 28s \\
13 & 8191 & 85 & 81 & 1.5m\\
13 & 8191 & 85 & 80 & 6.95m \\
13 & 8191 & 85 & 79 & 4.37h \\
13 & 8191 & 85 & 78 & 8.04h \\
13 & 8191 & 85 & 77 & 13.24h \\
14 & 16383 & 125 & 124 & 44s \\
14 & 16383 & 125 & 123 & 1.16m \\
14 & 16383 & 125 & 122 & 4.70m \\
14 & 16383 & 125 & 121 & 4.72m \\
14 & 16383 & 125 & 120 & 5.30m \\
14 & 16383 & 125 & 119 & 14.15m \\
14 & 16383 & 125 & 118 & 20.26m \\
14 & 16383 & 125 & 117 & 20.37m \\
14 & 16383 & 125 & 116 & 1.49h \\
14 & 16383 & 125 & 115 & 1.49h \\
15 & 32767 & 175 & 174 & 47.27m \\
15 & 32767 & 175 & 173 & 47.28m \\
15 & 32767 & 175 & 172 & 3.09h \\
15 & 32767 & 175 & 171 & 3.10h \\
16 & 65535 & 258 & 257 & 9.42m \\
16 & 65535 & 258 & 256 & 22.79m \\
16 & 65535 & 258 & 255 & 22.80m \\
16 & 65535 & 258 & 254 & 22.81m \\
17 & 131071 & 363 & 362 & 2.95D \\
17 & 131071 & 363 & 361 & 2.95D \\
17 & 131071 & 363 & 360 & 2.95D \\
\midrule
\showrowcolors
\bottomrule
\end{tabular}
\end{center}
\end{table}

\subsection{Using $\mathscr{A}$ as an m-sequences extension}

In this section, the following procedure is proposed:

\begin{itemize}
\item{Choose a primitive polynomial $f$ over $F_{2^m}$}
\item{Fix an initial element $a$ over $F_{2^m}$}
\item{Convert $f$ to a linear-feedback shift register $\mathscr{L}$}
\item{Expand the $\mathscr{L}$ to a binary sequence $L$, $\lvert L \rvert = 2^m-1$.}
\item{Launch $\mathscr{A}$ with $L$ as an input}
\end{itemize}

The primitive polynomials over $F_{2^m}$ could be calculated in advance. Furthermore, the PSL of $L$, where $L$ is seeded by some initial element $a$ over $F_{2^m}$, could be specially chosen to have the minimum possible value. This is easily achievable by using the following theorems (the proofs could be found in \cite{dimitrov2020aperiodic}): 

\begin{theorem}
\label{theorem1}
Given a binary sequence $B=b_0b_1 \cdots b_{n-1}$ with length $n$, the following property holds: $$\hat{C}_i(B \leftarrow 1) - \hat{C}_i(B) = b_0\left(b_{i+1} - b_{n-i-1}\right)$$
\end{theorem}

\begin{theorem}
\label{theorem2}
Given a binary sequence $B=b_0b_1 \cdots b_{n-1}$ with length $n$, the difference $\hat{C}_i(B \leftarrow \rho) - \hat{C}_i(B \leftarrow (\rho-1))$ is equal to $b_{(\rho-1) \bmod n}(b_{(i+\rho) \bmod n} - b_{(n-i+\rho-2) \bmod n})$.
\end{theorem}

The aforementioned procedure could be better illustrated by an example. If we fix $m=17$, we could pick the primitive polynomial $f = x^{17}+x^{14}+x^{12}+x^{10}+x^{9}+x+1$ over $F_{2^{17}}$.  Before converting $f$ to a linear-feedback shift register $\mathscr{L}$, we should fix the starting state of $\mathscr{L}$. Throughout this example, $a$ is fixed to the initial state of $\mathscr{L}$: $$\left[0,0,0,0,0,0,0,0,0,0,0,0,0,0,0,0,0,0,0,1,1\right]$$ Then, $\mathscr{L}$ is expanded to $L$. By using single instruction, multiple data (SIMD) capable device and starting with $L$, we could efficiently enumerate all $2^{17}$ different binary sequences generated by all possible starting states, to find the one generating the minimum PSL value. More formally, a value $\rho_{max}$, s.t. $ \forall\rho: {(L \leftarrow \rho_{max})}_{PSL} \leq {(L \leftarrow \rho)}_{PSL}$. Considering $f$ and the fixed value of $a$, in this specific case the value of $\rho_{max}$ is 15150, or more precisely, ${(L \leftarrow \rho_{max})}_{PSL} = 363$.

Experiments with initializing $\mathscr{A}$ ($\alpha$=6) with ${L \leftarrow \rho_{max}}$, instead of pseudo-randomly generated binary sequences, were made. We were able to repeatedly reach record-breaking binary sequences of length $131071$ having PSL equal to 359. The time required was less than 2 minutes, which was significant improvement over the time required for $\mathscr{A}$ (starting from pseudo-randomly generated sequences) to reach binary sequences with PSL close to 359: approximately 3 days. Leaving $\mathscr{A}$ to work for another 46 minutes it even reached binary sequences of length $131071$ with PSL 356.

The proposed procedure, as demonstrated, is highly efficient and is capable to reach binary sequences with $\mathscr{A}$-long lengths and record-breaking PSL values for few minutes. Unfortunately, it is applicable on binary sequences with lengths of the form $2^n-1$ only. However, throughout the next section, we provide another procedure which is able to generate binary sequences with length $p$ and record-breaking PSL values, where $p$ is a prime number. 

\subsection{Using $\mathscr{A}$ as an Legendre-sequences extension}

In this section, the following procedure is proposed:

\begin{itemize}
\item{Choose a prime number $p$}
\item{Generate the sequence $L=\left[t_1, t_2, \cdots, t_p \right]$}
\item{For $i$, s.t. $i \in N$, $1 \leq i \leq p$, and in case $i$ is a quadratic residue mod $p$, replace $t_i$ with 1. Otherwise, replace $t_i$ with -1.}
\item{Launch $\mathscr{A}$ with $L$ as an input}
\end{itemize}

As the numerical experiments suggested in \cite{dimitrov2020aperiodic}, it is highly unlikely that a Legendre sequence with length $p$, for $p>235723$, or any rotation of it, would yield a PSL value less than $\sqrt{p}$. Having this in mind, experiments with initializing $\mathscr{A}$ ($\alpha$=8) with a rotation of Legendre sequence with length 235747 were made (the next prime number after 235723). Again, by using SIMD capable devices, we have extracted the PSL-optimal rotation among all possible rotations of a  Legendre sequence with length 235747. More precisely, on rotation 60547, a binary sequences with PSL equal to 508 was yielded. Surprisingly, $\mathscr{A}$ was able to significantly optimize this binary sequence. As shown in Table \ref{tab:LegOptimized}, for less than 25 minutes, using only 1 thread of a Xeon-2640 CPU with a base frequency of 2.50 GHz, a binary sequences with PSL equal to 408 was found.

\begin{table}
\begin{center}
\caption{Time required for $\mathscr{A}$ to reach smaller PSL values, when launched from a rotated Legendre sequence with length 235747 and rotation value 60547.}
\label{tab:LegOptimized}
\ttfamily
\rowcolors{2}{light-gray}{light-cyan}
\begin{tabular}{lllll}
$PSL$ & $T$ \\
\toprule
496 & 1s\\
482 & 6s\\
462 & 15s\\
442 & 24s\\
422 & 9.75m\\
411 & 12.4m\\
410 & 18.9m\\
409 & 23.4m\\
408 & 23.7m\\

\midrule
\showrowcolors
\bottomrule
\end{tabular}
\end{center}
\end{table}

Since $\sqrt{235747} \approx 485.54$, it follows that 408 is significantly smaller than the expected value of 485.54. In fact, for leaving $\mathscr{A}$ for a total of  2.21 hours, a binary sequence with length 235747 and PSL 400, or 108$\blacktriangledown$, was reached: an example of such binary sequence is provided within the complimentary files. 

\section{Conclusions}
In this work, hybrid strategies for constructing binary sequences with optimal and near-optimal PSL values are suggested. By using the observations made throughout this paper, we were able to reveal binary sequences of almost any length with better PSL values than those known to the literature. As demonstrated, the proposed algorithms are applicable to short, mediocre, long or extremely long binary sequences.

 The number $\sqrt{n}$ appears frequently throughout the radar literature as an expected approximation of the optimal PSL value, among all binary sequences of length $n$. However, our numerical experiments suggest that $\sqrt{n}$ is a very pessimistic expectation.

\bibliographystyle{IEEEtran}
\bibliography{refs}

\end{document}